\documentclass[twocolumn,showpacs,preprintnumbers,amsmath,amssymb,subeqn,superscriptaddress]{revtex4}

\usepackage{graphicx}
\usepackage{dcolumn}
\usepackage{bm}
\usepackage{hyperref}  
\usepackage{latexsym}
\usepackage{color}
 
\begin{document}

\title{Canalizing Kauffman networks: 
Non-ergodicity and its effect on their critical behavior }

\author{Andr\'{e} Auto \surname{Moreira}} 
\email{auto@northwestern.edu}
\affiliation{ 
Department of Chemical and Biological Engineering,
Northwestern University, Evanston, IL 60208, USA}

\author{Lu\'{\i}s A. Nunes \surname{Amaral}} 
\email{amaral@northwestern.edu}
\affiliation{ 
Department of Chemical and Biological Engineering,
Northwestern University, Evanston, IL 60208, USA}

\begin{abstract} 
Boolean Networks have been used to study numerous phenomena,
including gene regulation, neural networks, social
interactions, and biological evolution. Here, we propose a
general method for determining the critical behavior of
Boolean systems built from arbitrary ensembles of Boolean
functions.  In particular, we solve the critical condition
for systems of units operating according to canalizing
functions and present strong numerical evidence that our
approach correctly predicts the phase transition from order
to chaos in such systems.
\end{abstract}

\pacs{64.60.Cn, 05.45.-a, 89.75.Hc}

\maketitle


Biological and social systems typically comprise a large
number of interacting units coupled through a nontrivial
network of interactions. Examples of such systems include
the metabolic processes in living cells~\cite{lee02} and
social interactions in human
groups~\cite{wassermann94,moreira04}. Remarkably, these
systems exhibit a high degree of self-organization that
ensures their continued functioning and allows them to
respond to environmental changes.  A challenging aspect in
the study of complex systems is how to model both the
diversity of the evolving units and the intricate structure
of their interactions~\cite{amaral04a}.

Discrete (agent-based) models are among the most common
methods used to tackle this challenge.  In particular,
Boolean networks~\cite{kauffman} (BNs) have been used to
model systems as varied as gene regulation
networks~\cite{kauffman}, evolution~\cite{stern99}, and
neuronal networks~\cite{kurten88}---see~\cite{aldana03} for a
review of BN and their applications. It has been shown that
BNs share many common properties with real
systems~\cite{kauffman,amaral04b}, the most remarkable
probably being a transition from an ordered to a chaotic
phase.

A BN consists of $N$ interacting units whose states
$\sigma_i$ are binary variables. Each unit $i$ is connected
to $k_i$ other units and its state is updated according to
a specific rule
\begin{equation}
\sigma_i(t+1)=F_i[\sigma_{i_1}(t),\sigma_{i_2}(t),...,\sigma_{i_{k_i}}(t)],
\label{e.rule}
\end{equation}
where $F_i$ is a Boolean function,
and the $\{\sigma_{i_j}\}$ are the states of the units
connected to $i$, which may or may not include $i$ itself.
Boolean functions are represented by a truth table that
lists the output of the function for each of the possible
set of input values.
For a function with $k$ variables there are $2^k$ possible
input set, yielding $2^{2^k}$ different possible functions.

The ensemble of functions ${\cal E}$ defines the probability
with which each function appears in the system.  In the
original formulation, BNs have the coupling connections
chosen at random and the Boolean functions $F_i$ drawn from
an ensemble ${\cal{E}}_{rand}(\rho)$, where $\rho$ is the
fraction of active states in the output of the functions. In
the following we will refer to $\rho$ as the ``bias'',
although the case $\rho=0.5$ is actually unbiased.  This
instance of the model is usually denoted Kauffman networks
or random Boolean networks (RBNs).

Typically, BNs display a transition from order to chaos.  In
the ordered phase, the network evolves toward limiting
cycles and, upon a perturbation, the system usually
converges back to the initial limiting cycle. In the chaotic
phase, the lengths of the attractor cycles grow
exponentially with $N$ and almost any perturbation will
drive the system toward a different attractor. The critical
behavior of RBNs has been determined by means of several
different
techniques~\cite{derrida86,luque00}. Not much,
however, is known about the critical behavior of BN with
other ensembles of functions.

In a recent paper, Shmulevich and
Kauffman~\cite{shmulevich04} suggested that the dynamical
behavior of a BN can be related to the ``average influence''
of the variables of its Boolean functions.
Here, we use the concept of damage spreading to demonstrate
the role of the influence in the dynamical behavior of BNs.
We show that, since BNs are nonlinear models not likely to
have ergodic dynamics, a naive average of the influence over
the whole phase-space of BNs does not necessarily yield a
correct estimate of the effective influence of the Boolean
variables.  We thus revise the definition of average
influence in order to account for the non-ergodicity of the
dynamics of BNs. Our definition enables us to derive the
critical condition of networks of canalizing Boolean
functions, a case of particular biological
relevance~\cite{harris01,kauffman03}. Finally, we show numerical
evidence that our method correctly predicts the critical
condition for networks of canalizing Boolean functions.


\begin{figure}
\begin{center}
\includegraphics[width=7cm]{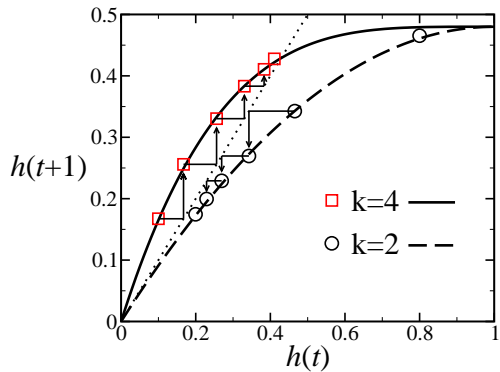}
\caption{
\baselineskip=12pt
The critical condition for Boolean networks (BN).  The
Hamming distance $h(t)$ is defined as the normalized
difference between two replicas of a BN at time $t$.  We
show in this figure the iterative mapping for the Hamming
distance for the case of random BNs with bias
$\rho=0.6$~\cite{derrida86}.  The solid line corresponds to
a network with  connectivity $k=4$ and the dashed line
to $k=2$. The symbols indicate the Hamming distances
obtained numerically for a few time steps in the evolution
of two biased networks with $k=4$ and $k=2$. The identity
mapping, indicated by the dotted line, and the arrows are
included to illustrate the time evolution of $h(t)$.  For
the case $k=4$, the Hamming distance remains at finite
value, while for $k=2$ it tends to zero. The system is in
the ordered phase when the Hamming distance converges to
zero.  Note that, $h=0$ is a fixed point of the mapping, but
is unstable whenever the iterative mapping at the origin
grows with a steeper slope than the identity line; cf.
Eq.~(\ref{e.crit}).
}
\label{f.hamming}
\end{center}
\end{figure}

The dynamics of BNs can be quantified by measuring the
spread of ``damage'' through the network.  This is done by
comparing the parallel evolution of two ``replicas'' of the
system. The replicas have identical Boolean functions and
coupling connections, but the initial state of the units in
the replicas differs in only a small fraction of the
units. The damage, which is also known as the Hamming
distance $h(t)$, is defined as the fraction of units that
are in different states in the two replicas. If, after some
transient time, the evolving replicas are likely to converge
to the same state, i.e., $h(t)\to{0}$, then the dynamics of
the system is robust with regard to small perturbations, a
signature of the ordered phase.  If, however, the replicas
are likely to never converge, then the dynamics is sensitive
to small perturbations to the initial state, a signature of
the chaotic phase. As discussed in the caption of
Fig.~\ref{f.hamming}, a system is in the ordered phase
whenever
\begin{equation}
\left. \frac{dh(t+1)}{dh(t)} \right|_{h(t) \to 0}<1,
\label{e.crit}
\end{equation}

Significantly, the susceptibility of unit $i$ to ``damage''
in its neighbors can be related to the influence of their
variables on $F_i$.  One defines the influence $I_j(F_i)$ of
the $j$th variable of a function $F_i$ as the probability
that the function $F_i$ changes its value when the value of
$\sigma_j$ is changed~\cite{kahn88,lyap}.
The average influence of a function
$I(F_i)\equiv\frac{1}{k_i}\sum_j{I_j(F_i)}$, and the average
influence $I({\cal E})$ of an ensemble ${\cal E}$ of Boolean
functions is $I({\cal
E})\equiv\langle{I(F_i)}\rangle_{\cal{E}}$, where
$\langle{...}\rangle_{\cal E}$ indicates an average over the
ensemble ${\cal{E}}$.
 
One can generalize this definition to multiple
variables~\cite{kahn88}: $I^{(1)} \equiv I$ is the average
influence of one variable, $I^{(2)}$ is the average
influence of two variables, and so on.   The probability
that an arbitrary unit is damaged in the next step depends
on the number $k_d$ of damaged inputs it gets and on the
influence $I^{(k_d)}$ of $k_d$ variables. Since the inputs
are an arbitrary sample of the entire network we can assume
that $k_d$ follows a binomial distribution and write the
evolution of the Hamming distance as 
\begin{equation}
h(t+1)= \sum_{k_d=1}^{k}{I^{(k_d)} \left(k \atop k_d \right) 
[h(t)]^{k_d} [1-h(t)]^{k-k_d}}
\label{e.hamming2}
\end{equation}
where $\left( k \atop i \right)$ is the binomial
coefficient.    Thus, the influences $I^{(k_d)}$
determine the shape of the iterative mapping of $h(t)$.
Inserting Eq.~(\ref{e.hamming2}) into Eq.~(\ref{e.crit}), we
have that the critical condition depends only on the average
influence of one variable, 
\begin{equation}
I({\cal{E}},k_c) k_c=1.
\label{e.kcsc}
\end{equation}
Equation~(\ref{e.kcsc}) enables us to determine the critical
condition for BNs with arbitrary ensembles of
functions~\cite{sdk}.

This is not trivial, however. The difficulty in using
Eq.~(\ref{e.kcsc}) lies in computing the influence of the
variables of the Boolean functions  present in the
network. In principle, the influence of the variables can
be determined by counting in the truth table the number of
times that by changing the value of only one variable
results in a change in the value of $F$.
This approach, which was explored in~\cite{shmulevich04},
implicitly assumes ergodicity, that is, all inputs can arise
with the same probability during evolution, and time average
over the states visited by the network yields the same
result as average over the whole phase-space. This is an
implausible assumption which is unlikely to hold for the
dynamics of arbitrary BNs.

In some instances, however, an equaly weighted average does yield to
correct results.
An example is the ensamble of RBNs~\cite{luque00}. Note
that this does not imply that RBNs are ergodic. In fact, the
dynamics of BNs in general converge to limiting cycles that
occupy only a fraction of the entire phase-space. To
correctly average the influence of the Boolean variables,
one must measure the influence only on those states
composing the limiting cycles.

We can verify in which cases an equally weighted average can
work.  If one assumes that the states of the neighbors of a
unit are not correlated with the state of the unit it self
(random-graph approximation), it follows that the input
acting on the unit is a statistical sample of the whole
network. Thus, the probability of a certain input depends
on the fraction $q$ of units that are in the active
state. That is, if the network has a bias toward activity,
$q>0.5$, the inputs with more 1s will be more frequent than
the inputs with more 0s. Therefore, the activity $q$ of the
network should be taken into account when computing the
average influence of the BN. The reason why a simple average
over the whole phase-space works in RBNs and a few other
ensambles is that, on these networks, the influence does not
depend on $q$, thus, averaging over the states of the
limiting cycles yields the same result as averaging over the
whole phase-space. As we will demonstrate later, this
property does not generally hold for arbitrary ensembles of
Boolean functions.

In the following, we focus on the ensemble of canalizing
Boolean functions ${\cal E}_{can}$. Studies of gene
regulation in eukaryots have showed that the Boolean
idealization is a good approximation for the non-linear
dynamics of this system and that the gene regulating
mechanisms have a strong bias toward canalizing
functions~\cite{harris01}.  A Boolean function is canalizing
if whenever one variable, the canalizing variable, takes a
given value, the canalizing value, the function always
yields the same output. The ensemble of canalizing functions
can be separated into four mutually exclusive classes of
functions:
\begin{subequations}
\begin{eqnarray}
F(\sigma_1,\sigma_2,...)=&
\makebox[1.1cm][r]{ }~\makebox[0.5cm][l]{$\displaystyle \sigma_1$}
~\makebox[0.9cm][c]{OR}&G(\sigma_2,...),
\label{e.cana1} \\
F(\sigma_1,\sigma_2,...)=&
\makebox[1.1cm][r]{$\displaystyle ($NOT}~\makebox[0.5cm][l]{$\displaystyle \sigma_1)$}
~\makebox[0.9cm][c]{AND}&G(\sigma_2,...),
\label{e.cana2} \\
F(\sigma_1,\sigma_2,...)=&
\makebox[1.1cm][r]{$\displaystyle ($NOT}~\makebox[0.5cm][l]{$\displaystyle \sigma_1)$}
~\makebox[0.9cm][c]{OR}&G(\sigma_2,...),
\label{e.cana3} \\
F(\sigma_1,\sigma_2,...)=&
\makebox[1.1cm][r]{ }~\makebox[0.5cm][l]{$\displaystyle \sigma_1$}
~\makebox[0.9cm][c]{AND}&G(\sigma_2,...),
\label{e.cana4} 
\end{eqnarray}
\label{eqs}
\end{subequations} \noindent
where $\sigma_1$ is the canalizing variable, ``AND,''
``OR,'' and ``NOT'' are the logical Boolean operators, and
$G$ is the non-canalizing part of the function that carries
the dependence on the remaining variables. Each of this
classes represents a different type of regulation. The class
described by~(\ref{e.cana1}) represents ``sufficient
activators,'' that is, $\sigma_1=1$ is sufficient to assure
an active state for the unit. The class described
by~(\ref{e.cana2}) represents ``sufficient repressors,''
that is, $\sigma_1=1$ always results in an inactive state
for the unit. The classes described
by~(\ref{e.cana3})~and~(\ref{e.cana4}) represent ``necessary
repressors'' and ``necessary activators,'' respectively. In
these cases, $\sigma_1=0$ is also enough to determine the
output of the function.

\begin{figure}
\begin{center}
\includegraphics[width=7.8cm]{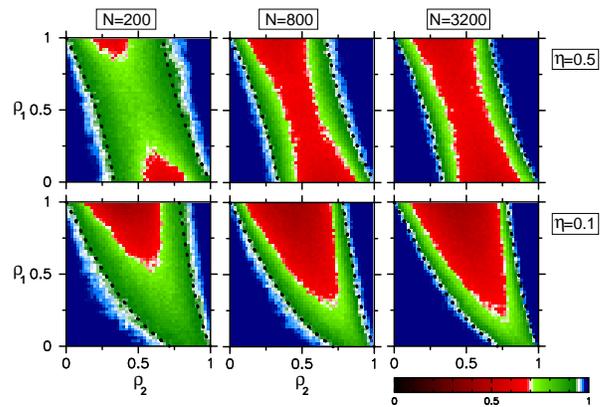}
\caption{
\baselineskip=12pt
Order-chaos phase-transition for networks of canalizing
Boolean functions. The color coding represents the
probability that, after changing the state of a random unit
of a network in a limiting cycle, the system returns to the
same cycle. The resilience of the network to small
perturbations is a signature of the ordered phase.   
 The
parameter $\eta$ gives the probability with which a Boolean
function in the network is canalized by an active input.  In
the trivial case $\eta=0.5$, the average influence does not
depends on the activity of the network and the influence can
be computed with an average over the truth table of the
Boolean functions in the network, and one finds the critical
condition when $kI=1$ in Eq.~(\ref{e.half}).  In the
non-trivial case $\eta=0.1$ one finds no difference in the
value of the influence when computed from the truth table of
the functions. However, it is clear that the critical curve
is sensible to the value of $\eta$. The difference is due to
the fact that in the latter case the average influence
depends on the activity of the network and one has to make a
weighted average over the states occupied by the limiting
cycles and the critical condition is $kI=1$ in
Eq.~(\ref{e.nonerg}).  In both cases, as the network grows,
the transition becomes sharper and approaches the critical
curves (dotted lines).} 
\label{f.quench} 
\end{center}
\end{figure}

  The average influence for the ensamble ${\cal{E}}_{can}$
depends on the probability $P_{can}$ with which $\sigma_1$
takes the canalizing value.  For classes described
by~(\ref{e.cana1})~and~(\ref{e.cana2}), $\sigma_1$ gives a
sufficient condition for activation or repression
respectively. This means that for these classes the
canalizing value is an active state. On other hand, for
classes~(\ref{e.cana3})~and~(\ref{e.cana4}) the canalizing
value is an inactive state. If both cases are
equally present on the network, one has always
$P_{can}=0.5$. However, if  one of the canalizing values
is more frequent than the other, $P_{can}$ will depend on the
fraction $q$ of units in the active state. To account for
this effect we define $\eta$ as the fraction of the
functions in the ensemble that fall into classes
(\ref{e.cana1})~and~(\ref{e.cana2}). Thus, the probability
that the canalizing variable takes the canalizing value is
\begin{equation} 
P_{can}=q\eta+(1-q)(1-\eta).
\label{e.pcan}
\end{equation}

 The next step is determining the average 
activity $q$ of the network when the limiting cycles are
reached. To do this, we need to define some relevant
parameters characterizing the ensemble of canalizing
functions.  Note that, for the classes described by
Eqs.~(\ref{e.cana1})~and~(\ref{e.cana3}), the use of the
``OR'' operator means that the values of $\sigma_1$ and $G$
give two alternative conditions yielding an active output,
while for the classes described by
Eqs.~(\ref{e.cana2})~and~(\ref{e.cana4}), the ``AND''
operator means that the values of $\sigma_1$ and $G$ give
two necessary conditions for obtaining an active output. The
use of the ``OR'' operator thus results in a bias toward
activity. To quantify this bias, we define $\rho_1$ as the
fraction of the functions in the ensemble that fall into
classes~(\ref{e.cana1})~and~(\ref{e.cana3}). Note that, the
bias of the canalizing functions toward the active state
will also depend on $G$, the non-canalizing part of
$F$. We assume that $G$ is chosen as a random Boolean
function with bias $\rho_2$.

  It is possible to measure the probability that a random
input results in an active output.  This probability is the
average bias $\rho_{can}=(\rho_1+\rho_2)/2$ of the ensemble
of canalizing functions ${\cal{E}}_{can}$. However, one can 
not assume that in the limiting cycles any input happen with
the same chance. For the ensemble ${\cal{E}}_{can}$, 
the average activity for the limiting cycles is given by
\begin{equation}
q=\rho_1P_{can}+\rho_2(1-P_{can}),
\label{e.activity}
\end{equation}
where the first term on the right accounts for the
probability that the function is being canalized to activity
and the second for the probability that the function $G$ is
driving the function to activity.

We can now proceed and calculate the average influence for
the ensemble of canalizing functions. We will consider first
only the average influence of the canalizing variable $I_1$,
which is given by the probability that $G=0$ when the OR
operator is chosen, plus the probability that $G=1$ when the
AND operator is chosen;
$I_1=\rho_1(1-\rho_2)+(1-\rho_1)\rho_2$. The influence of
the remaining variables $I_i$ depends on the probability
that the functions are not locked by the canalizing variable,
$1-P_{can}$, and on the bias $\rho_2$ of $G$. Finally, we
have $I_i=2\rho_2(1-\rho_2)(1-P_{can})$, and:
\begin{equation}
k I= \rho_1+\rho_2-2\rho_1\rho_2+2\rho_2(1-\rho_2)[\eta+q(1-2\eta)](k-1).
\label{e.nonerg}
\end{equation}

If one assumes that all inputs in the truth table
contribute with the same weight to the average, then
$P_{can}=0.5$, and
\begin{equation}
k I=\rho_{1}+\rho_{2}-2\rho_{1}\rho_{2}+(k-1)\rho_{2}(1-\rho_{2}).
\label{e.half}
\end{equation}
One of the cases where Eq.~\ref{e.half} works is when $\eta=0.5$.

We next test our theoretical results against numerical
simulations of BNs of canalizing functions. This is done by
building random networks with Boolean functions obeying the
ensemble of canalizing functions described by
Eq.~(\ref{eqs}). We assign random initial states to the
networks and let them evolve until they reach a limiting
cycle~\cite{weight}. We then make a perturbation by changing
the state of one of the units in the network. The resilience
of the system to this ``damage'' is the probability that,
after the perturbation, the system converges back to the
initial limiting cycle~\cite{estim}.  
We show in Fig.~\ref{f.quench} that, as the system size
grows, the transition from order to chaos becomes sharper
and approaches a critical condition where $kI=1$;
cf. Eq.~(\ref{e.nonerg}).

Note that, when the network has a bias in the canalizing
value, $\eta=0.1$, there is a considerable reduction in the
region occupied by the chaotic phase, mainly in the region
where the network is biased to the inactive state: low
$\rho_1$ and $\rho_2$. This bias for an inactive canalizing
value was observed in the mechanisms of gene
regulation~\cite{kauffman03} where the transcription of a
given gene may depend on the presence of several activator
proteins, that is, a single inactive input---the absence of
the one of the activators---can result in an inactive
state---no transcription.


The major finding of this study is that, by using the
concepts of influence of Boolean variables and damage
spreading, we are able to obtain the critical behavior of
Boolean networks built from arbitrary ensembles of
functions. We show that for most networks the effective
influence of the variables cannot be obtained by a simple
average over the truth table of the functions. We further
obtain an expression for the influence of the variables for
networks of canalizing Boolean functions and present strong
numerical evidence that our method can accurately predict
the critical transition for these networks. Our work
suggests that the approach described here can solve the
critical transition of other ensembles of Boolean functions
such as nested canalizing
functions~\cite{kauffman03}---which are thought to be a
valuable model for the description of gene regulation
networks---or random threshold functions~\cite{kurten88}---a
common model for neural networks.

 
We thank A. D\'{\i}az-Guilera, L. Guzm\'an-Vargas,
D. B. Stouffer, M. Sales, and R. Guimer\`a for fruitful
discussions. LANA thanks the support of a Searle Leadership
Fund Award and a NIH/NIGMS K-25 award.



\begin{thebibliography}{99}
\bibitem{lyap} 
The concept of influence is related to Boolean derivatives
and Lyapunov exponents of BNs~\protect\cite{luque00} and to
the activity defined in~\protect\cite{shmulevich04}.
\bibitem{sdk} 
This critical condition can be generalized to the case where
the number of variables of the units is drawn from a
distribution $P(k)$, becoming
$\int{dk~P(k)~S[{\cal{E}}(k)]~k}=1$.
\bibitem{weight} In this way, the mean values we obtain 
represent a weighted average where each attractor is
weighted by the size of its basin of attraction.
\bibitem{estim}
Our estimate is obtained by averaging over $10^4$ different instances
of the network for each set of parameter values.
\end{thebibliography}

\end{document}